\documentclass[11pt,twoside]{article}
\usepackage{./asp2014}{}

\aspSuppressVolSlug
\resetcounters

\begin{document}

\title{Early B-type stars with resolved Zeeman split lines}

\author{S.~Hubrig$^1$, S.~P.~J\"arvinen$^1$, M.~Sch\"oller$^2$, J.~F. Gonz\'alez$^3$}
\affil{$^1$ Leibniz-Institut f\"ur Astrophysik Potsdam (AIP), An~der Sternwarte~16, 14482~Potsdam, Germany}
\affil{$^2$ European Southern Observatory, Karl-Schwarzschild-Str.~2, 85748~Garching, Germany}
\affil{$^3$ Instituto de Ciencias Astron\'omicas, de la Tierra y del Espacio (ICATE), 
Av.~Espana Sur~1512, CC~49, 5400~San~Juan, Argentina}

\paperauthor{S. Hubrig}{shubrig@aip.de}{}{Leibniz-Institute f\"ur Astrophysik Potsdam (AIP)}{}{Potsdam}{}{14482}{Germany} 
\paperauthor{S.~P. J\"arvinen}{}{}{Leibniz-Institute f\"ur Astrophysik Potsdam (AIP)}{}{Potsdam}{}{14482}{Germany}
\paperauthor{M. Sch\"oller}{}{}{European Southern Observatory}{}{Garching}{}{85748}{Germany}
\paperauthor{J.~F. Gonz\'alez}{}{}{Instituto de Ciencias Astron\'omicas (ICATE)}{}{San Juan}{}{5400}{Argentina}

\begin{abstract}

Almost three decades ago, \citet{Mathys1990} demonstrated the importance of studying Ap 
stars showing resolved Zeeman split Fe~{\sc ii}~6147.7 and 6149.2 lines. Such Zeeman split lines 
can be seen in stars whose projected rotational velocity is sufficiently small and 
whose magnetic field is strong enough to exceed the rotational Doppler broadening. 
Observations of resolved Zeeman split lines permit the diagnosis of the average of 
the modulus  of the magnetic field over the visible stellar hemisphere.
Although Zeeman splitting is not expected in faster rotating hot massive stars,
we have recently been discovering  early B-type stars displaying magnetically split spectral lines.
\end{abstract}

\section{Introduction}

A number of Ap  and late-type Bp stars with strong magnetic fields show resolved Zeeman 
split spectral lines,
allowing to set additional constraints on the magnetic field geometry by measuring the mean magnetic 
field modulus, i.e.\ the average over the  visible stellar hemisphere of the modulus of the magnetic field 
vector, weighted by the local line intensity, following the relation given
e.g.\ by \citet{Hubrig2007} and \citet{Mathys2017}.
Such Zeeman split lines were recently discovered in a few early B-type stars with low projected
rotational velocity and strong, kG order magnetic fields.
In Fig.~\ref{fig:reduced}, we present several examples of magnetically split lines in three
such stars, HD\,58260, HD\,96446, and HD\,149277. We briefly characterize them in the following sections.

\articlefigure[width=.75\textwidth]{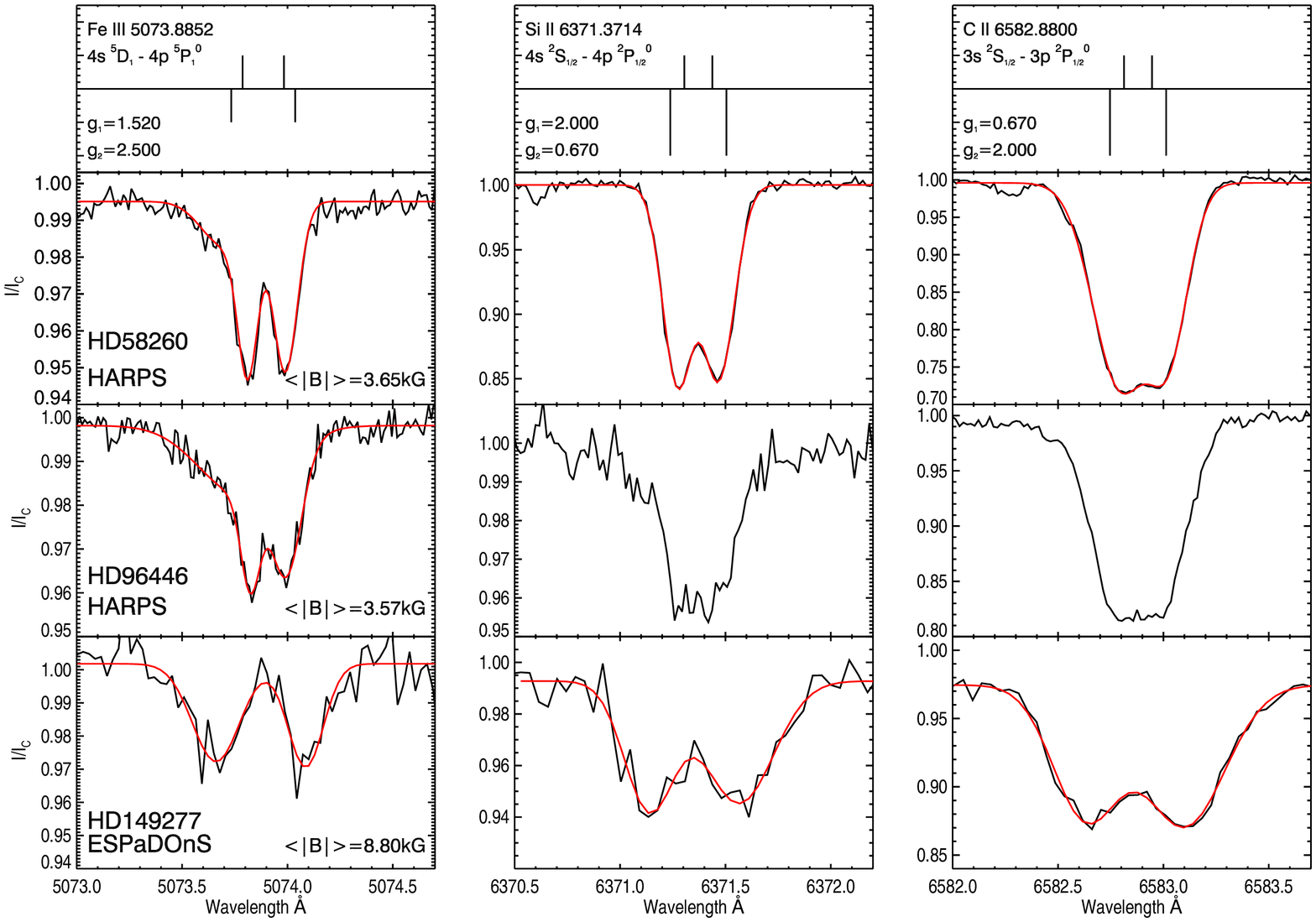}{fig:reduced}{
The magnetically split lines Fe\,{\sc iii}~$\lambda$5074, Si\,{\sc ii}~$\lambda$6371, and 
C\,{\sc ii}~$\lambda$6583 in 
high-resolution Stokes~$I$ spectra of the early B-type stars HD\,58260, HD\,96446, and HD\,149277.
The red lines denote the fit of a multi-Gaussian to the data.
For two lines in the spectrum of HD\,96446, the splitting is not sufficient to allow a proper fit.
}

\section{The B2Vp star HD\,58260 with a 3.65~kG mean magnetic field modulus}

According to 
\citet{Bohlender1987}, the longitudinal magnetic field of this star has been constant at 
2.3\,kG between 1977 and 1986. \citet{Shultz2018}  suggested a non-variability 
of the mean longitudinal magnetic field $\left<B_z\right>$ over a time scale of about 
35~years. Our measurements  presented in Fig.~\ref{fig:reduced2}
show $\left<B_z\right>$ values in the range 1.4--1.5\,kG, and are weaker than the values reported by 
\citet{Shultz2018} by $\sim$300\,G \citep{Jarvinen2018}.

\articlefigure[width=.5\textwidth]{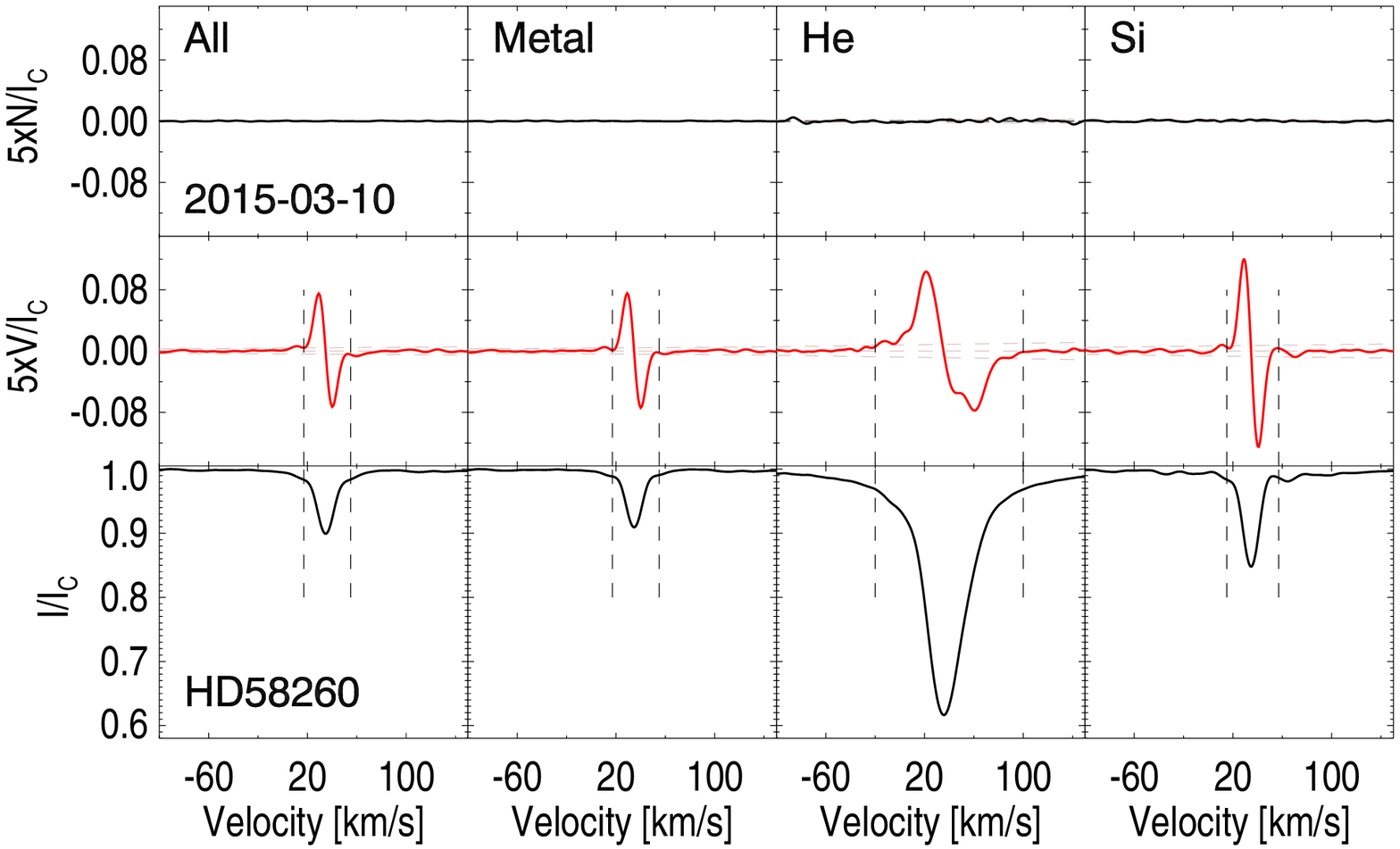}{fig:reduced2}{
LSD Stokes~$I$, Stokes~$V$, and diagnostic null $N$ profiles (from bottom to top) calculated for 
HD\,58260 using a line mask containing all spectral lines, a line mask with only metal 
lines, a line mask containing exclusively He\,{\sc i} lines, and a line mask containing exclusively 
Si\,{\sc iii} lines (from left to right). The vertical dashed lines indicate the integration ranges for the 
determination of  $\left<B_z\right>$.
The Stokes~$V$ and $N$ profiles were expanded by 
a factor of 5 and shifted upwards for better visibility. The red 
dashed lines indicate the standard deviations for the Stokes~$V$ and $N$ spectra.
}

\section{The B2Vp star HD\,96446 with a 3.9~kG mean magnetic field modulus}

A rotation period of 23.4\,days was recently identified using an extensive spectroscopical 
time series data set obtained by Gonz\'alez et al. (in preparation). \citet{Jarvinen2017} 
showed the strong impact of $\beta$ Cephei-like pulsations on the 
line profile and magnetic field measurements. The measurements carried out using the 
He and Si line masks separately revealed that the individual field strengths are differing 
at some rotation phases by about 1\,kG. These elements are usually distributed inhomogeneously 
on the stellar surface of He-rich stars. For the analysis of the magnetic field 
behaviour in the star HD\,96446 over the refined rotation period, we used a 
sample of archival HARPS\-pol observations completed by our own most recent observations. 
The phase distribution of the mean longitudinal magnetic field
measurements using three different line masks, including the best sinusoidal
fits, is presented in Fig.~\ref{fig:reduced3}, where we also show measurements
by \citet{neiner2012}, which were based on selected metallic lines. 
The difference between the measurements carried out for the line lists
containing He and Si separately shows that the corresponding field strengths
are differing at some rotation phases by about 1\,kG. The measurements based on the He lines show an
almost flat distribution.

\articlefigure[width=.6\textwidth]{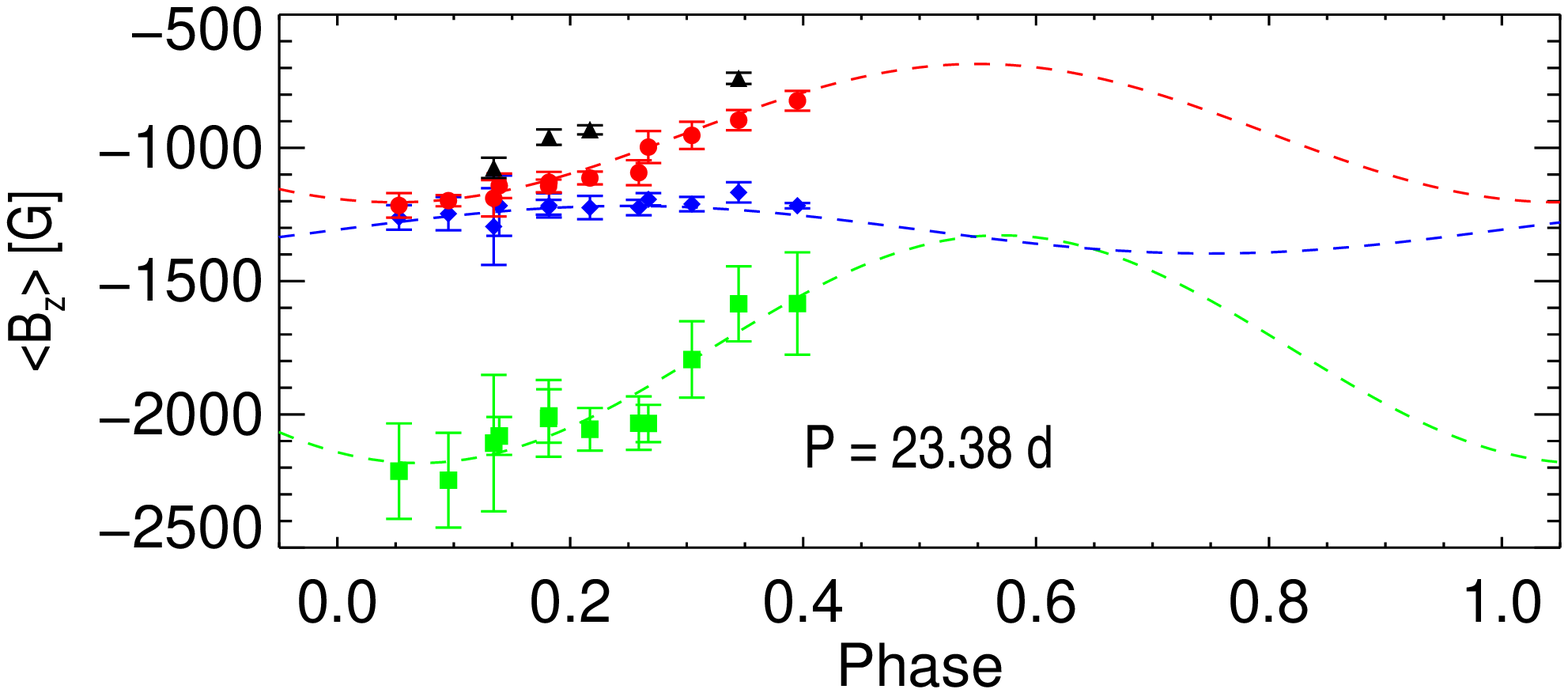}{fig:reduced3}{
Magnetic field measurement distribution for HD\,96446 over the rotational phase using
the ephemeris $T = 2\,455\,000.0 + 23\fd38 \ E$. Red filled circles
represent values measured from all lines, green squares are from Si lines,
and blue diamonds are from He lines. The dashed lines represent the respective fits.
In addition, the values measured by
\citet{neiner2012} using only metal lines are denoted with black triangles. 
}

\section{The B2IV/V star HD\,149277 with a mean magnetic field modulus of $\sim$9\,kG}

HD\,149277 is a rare SB2 system with a slowly rotating magnetic He-rich primary with 
$P_{\rm rot}=25.4$\,d \citep{Shultz2018}. Using  high-resolution polarimetric spectra of HD\,149277
acquired with ESPaDOnS at the Canada-France-Hawaii Telescope, \citet{Shultz2018} 
reported on the presence of a rather strong longitudinal magnetic field with 
$\left<B_{\rm z}\right>_{\rm max}=3.3\pm0.1$\,kG.
The rotation period of 25.4\,d was determined using mean longitudinal magnetic field measurements.
To estimate the orbital parameters of the SB2 system HD\,149277, we downloaded all 
publically available  ESPaDOnS spectra obtained from 2012 to 2016. 
These spectra revealed $P_{\rm orb}=11.5192 \pm 0.0005$\,d indicating strong subsynchronous 
rotation of the primary component \citep{Gonzalez2018}. Such a strong
subsynchronous rotation was not detected in any other SB2 system with a magnetic chemically 
peculiar component and is most likely caused by magnetic braking. 
Our inspection of the ESPaDOnS spectra revealed the presence of several resolved Zeeman split lines.
Their presence has not been reported previously.
In Fig.~\ref{fig:reduced4}, we present 
the rotational variability of the resolved Zeeman split Fe\,{\sc iii}~5073.6 line in the primary component.
Interestingly, as illustrated in this figure, we observe that in certain phases the Zeeman split
lines are asymmetric about the line centers with the blue
components appearing less deep than the red components, and with the blue components deeper 
in other phases. Such asymmetries occur in the same way in other transitions, indicating that 
the rotational Doppler effect is non-negligible, i.e.\
various parts of the stellar surface characterized by different magnetic field strengths contribute to the
line profile.

\articlefigure[width=0.9\columnwidth,height=8.1cm, bb=70 30 557 737]{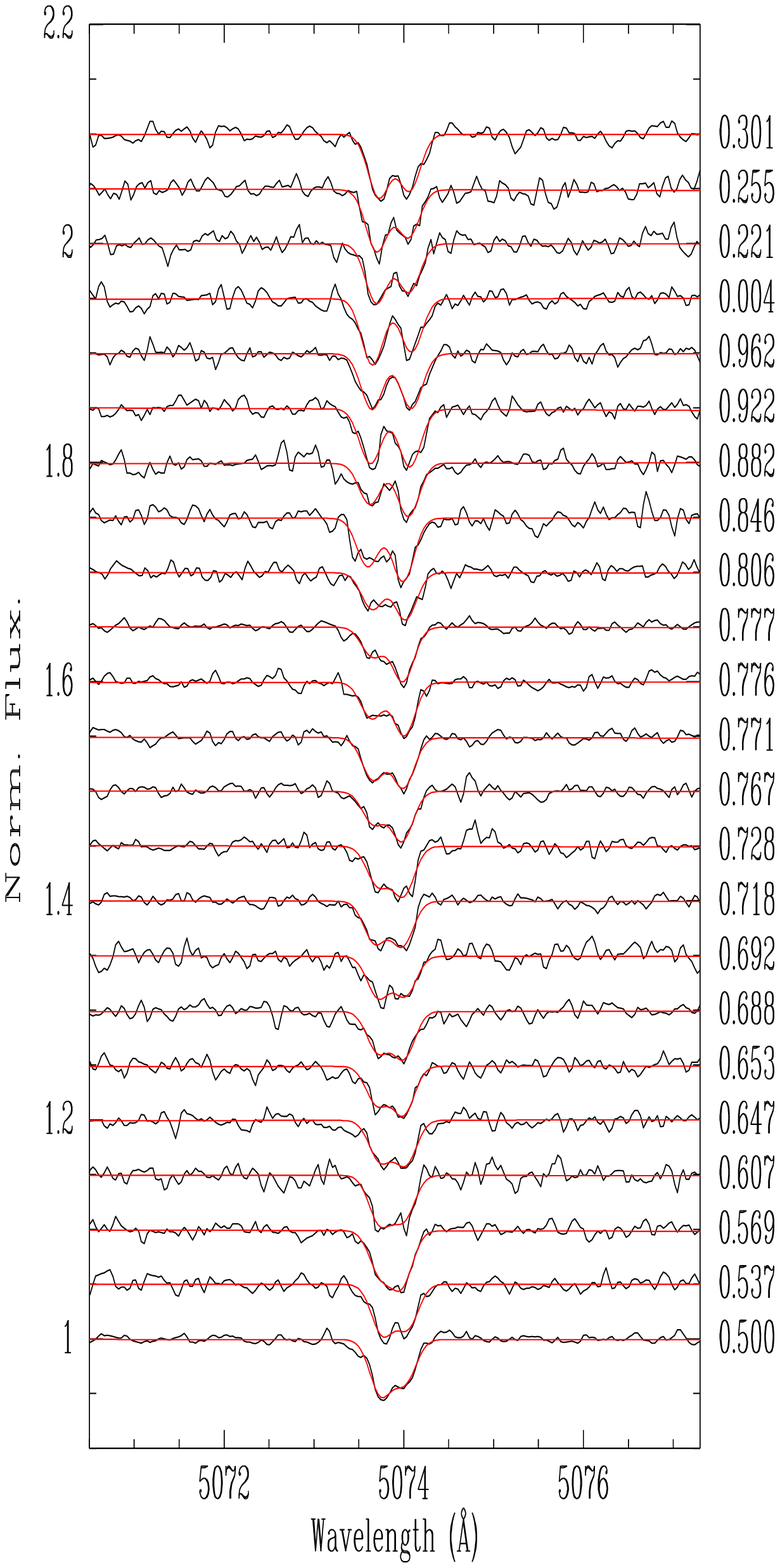}{fig:reduced4}{
Variability of the magnetically split Fe\,{\sc iii}~5073.6 line over the rotation period in 
the disentangled spectra of the primary component in the system HD\,149277.
Spectra are vertically offset for better visibility and are sorted in rotational phase
(see right side; starting at 0.5) from bottom to top.
The red lines represent a double-Gaussian fit to the data.
}

The maximum of the magnetic field modulus coincides roughly with
the positive extremum of the longitudinal magnetic field, whereas the
minimum of the modulus is coinciding with the negative extremum of the longitudinal magnetic field. 
No evidence for a longitudinal magnetic field was seen in the circularly polarized spectra of 
the secondary component.

\end{document}